\documentclass{sig-alternate}
\usepackage{comment}
\usepackage{color}
\usepackage{amsmath}
\usepackage{tikz}
\usepackage{tikzscale}
\usetikzlibrary{arrows,decorations.markings}
\usepackage[colorinlistoftodos]{todonotes}
\usepackage{booktabs}
\usepackage{colortbl}

\begin{document}
%
\conferenceinfo{BigCHat Workshop KDD}{'15 Sidney, Australia}

\title{A latent shared-component generative model for real-time disease surveillance using Twitter data
}
%
%
%
%
%

\numberofauthors{6} 
%
\author{
%
%
Roberto C.S.N.P. Souza, 
Denise E.F de Brito,
Renato M. Assunção,
Wagner Meira Jr. \\
       \affaddr{Department of Computer Science.}
       \affaddr{Universidade Federal de Minas Gerais, Brazil}\\
       \email{\{nalon,denisebrit,assuncao,meira\}@dcc.ufmg.br }
}

\maketitle
\begin{abstract}

Exploiting the large amount of available data for addressing relevant social
problems has been one of the key challenges in data mining. Such efforts have
been recently named ``data science for social good'' and attracted the attention
of several researchers and institutions.  We give a contribution in this
objective in this paper considering a difficult public health problem, the
timely monitoring of dengue epidemics in small geographical areas. We develop
a generative simple yet effective model to connect the fluctuations of
disease cases and disease-related Twitter posts. We considered a hidden Markov
process driving both, the fluctuations in dengue reported cases and the tweets
issued in each region. We add a stable but random source of tweets to represent
the posts when no disease cases are recorded.  The model is learned through a
Markov chain Monte Carlo algorithm that produces the posterior distribution of
the relevant parameters. Using data from a significant number of large
Brazilian towns, we demonstrate empirically that our model is able to predict
well the next weeks of the disease counts using the tweets and disease cases
jointly.

\end{abstract}

\category{H.2.8}{Database Applications}{Data mining}


\keywords{Data mining, Disease surveillance, Dengue, Social media data}

\section{Introduction}
\label{sec:intro}

Exploiting the large amount of available data for addressing relevant social
problems has been one of the key challenges in data mining. Such efforts have
been recently named ``data science for social good'' and attracted the attention
of several researchers and institutions.  One example of such scenario is
associated with the rise of text-based social media, which have created a whole
new ecosystem composed of individuals sharing daily live information. The
popularization of such platforms led people to report about almost everything,
from their common activities, to their social interactions, preferences and
health conditions \cite{Dredze}.  Such a large amount of user generated
content provides unprecedented valuable information that may be used for a
variety of purposes.  Among all the possibilities, event detection has received
significant attention.  In this context, data from social media channels are
leveraged in order to predict real-world events, such as natural disasters
\cite{Sakaki2010} and disease outbreaks \cite{Lampos}. The common approach in
this case is to consider users posting to a social media as sensors and their
respective messages about an event as an indicator of its occurrence/intensity.
In this sense, the intensity of an event is predicted as a function of the
number of messages related to that event, which should be specified by a set of
keywords.

In the particular case of disease surveillance, using data from social media
provides a great advantage over traditional systems~\cite{Prieto}.
Usually, such systems are almost entirely manual and require a set of
epidemiological procedures in order to monitor the intensity and location of
the disease spread.  The complete process may result in a long delay for
decision making and prevents early monitoring.  In a strikingly different way,
messages posted to social platforms can be collected and analyzed almost
instantly, enabling timely actions to control disease outbreaks or swift
geographical spread of cases.

Dengue is an infectious disease that is currently a major concern for Brazilian
public health officials (see section \ref{sec:dengue}).  There is still no
effective cure or prevention (i.e., vaccine) for dengue fever and its painful
symptoms and occasional associated death make it a common daily conversation
topic. During the epidemic periods, it becomes a much more common issue in
social interactions when people exchange their personal, close friends or
relatives experiences with the disease, as well as discussions about the public
health policies and counter measures. There is clear potential to use social
network posts to probe and to monitor the fluctuations on the disease incidence
rate, in addition to existing official disease surveillance systems, as
demonstrated in previous works (see Section~\ref{sec:relwork}). However, 
the models proposed so far are not accurate enough, considering the diversity
of factors that determine a dengue surge.

These are the main motivations behind our latent shared-component generative model to predict dengue
outbreaks in Brazilian urban areas using data collected from Twitter. In this
paper, we propose a concise yet flexible analytical model for the simultaneous
rise and fall patterns of the dengue cases and the dengue-related tweets.  We
envision a hierarchical and stochastic structure that is parsimonious in the
number of parameters. In our model, the disease rate of dengue cases is a
latent random process that drives the weekly reported number of dengue cases.
As it is common in epidemic modeling, to reflect the smoothly varying and
time-dependent aspect of the disease, we use a hidden Markov model for this
underlying rate \cite{Anderson1992, Matsubara2014, Bailey1975}.

The novelty in our model is the joint effect of this underlying process in the
dengue-related tweets time series. The weekly count of these tweets has two
independent components. One of them is a Poisson process issuing a stable and
random amount of tweets reflecting the relevance of the dengue in the usual
conversation in Brazil. The evidence for the presence of this component has
been verified in our dataset by observing a certain number of successive weeks
with zero dengue cases but with a relatively small number of tweets mentioning
the disease. The motivation for our second component in the tweets time series
is the clear pattern in the observed  data that more prominent temporal
increase or decrease in the disease counts is accompanied by similar movements
in the tweets time series.

In section \ref{sec:method}, we explain how we connect these two components to
model the tweets and, at the same time, the disease cases. We show the results
of the application of this model to the analysis of dengue in the largest
Brazilian towns. However, we consider that the general methodological framework
and the main considerations are widely applicable, not only to other diseases,
but also to other scenarios.

\subsection{Dengue overview}
\label{sec:dengue}

With an estimated 50-100 million infections globally per year
\cite{nature2013}, dengue is currently regarded as the most important
mosquito-borne viral disease. Dengue affects over 100 endemic countries in
tropical and sub-tropical regions of the world, mostly in Asia, the Pacific
Region and the Americas.  Presenting four distinct viral serotypes, dengue
fever may range from severe flu-like illness up to a potentially lethal
complication known as severe (or hemorrhagic) dengue. The World Health
Organization estimates that 3.9 billion people are at risk of infection with
dengue viruses. However, the true impact of the disease is, sometimes,
difficult to assess due to misdiagnosis and underreporting~\cite{who}. Global
dengue incidence still grows in number/severity of cases and also in the amount
of new affected areas. This is most due to modern climate changes,
socioeconomics and viral evolution~\cite{clinicaldengue2013}. However, the
potential drivers of dengue are often difficult to detect and factor out. Since
there is no current approved vaccine to protect the population against the
virus \cite{clinicaldengue2013}, epidemiological surveillance and effective vector control are
still the mainstay of dengue prevention.

The Brazilian traditional surveillance system is almost entirely manual and
relies on the ability to observe early cases of dengue for each location and time
period.  This process usually results in a huge delay for data acquisition.  In
2013, the Brazilian Ministry of Health reported almost 1.5 million cases of
dengue infection, resulting in 674 deaths \cite{portalsaude}. The earlier an
outbreak is detected, the greater is the opportunity for effective
intervention.  In this sense, social media provides a real time source of
information to monitor and predict such outbreaks.  The model proposed in this
work leverages social media data in order to provide predictive estimations of
dengue incidence allowing timely actions.

\section{Related work}
\label{sec:relwork}

In recent years, a large number of researches focused on leveraging data from
text-based social media towards monitoring and predicting real-world
events~\cite{Sakaki2010, STSS, Lampos}.  Among all types of events considered,
those related to disease surveillance usually receive significant attention,
due to their potential impact on life conditions. Several studies show that the
wealth of user generated content available through social media channels is a
rich source of information to track the spread of a wide range of diseases,
therefore enabling timely supporting actions.

In \cite{Signorini} tweets collected using a set of keywords related to the
H1N1 virus were used to monitor the public concern about the disease in the
United States. Flu was the major concern in \cite{Lampos, Culotta2010},
where regression models were applied to estimate the disease incidence rate
using data from Twitter and comparing to the reports provided by the official
organizations. Regression models were also employed in a more recent work
\cite{Santos}, combining data from Twitter and search engines to estimate the
flu incidence in Portugal.

Despite flu being the most common disease monitored from social media, other
studies employed this approach to the surveillance of other diseases. For
instance, in \cite{Chunara} data from Twitter was used to estimate the
incidence of cholera in Haiti, collecting the messages containing the terms
``(\#)cholera''.  In a more comprehensive perspective \cite{Dredze, Prieto}
considered a broader range of public health dimensions, expanding the
analysis for different ailments, such as depression, obesity and others,
showing that social media has a wide applicability for public health research.
 
Most of the aforementioned works focused on the disease surveillance at a
global scale in the sense that one performs the analysis at state, region or
even national level. Such coarse grain analysis lacks the specificity necessary
for supporting epidemics counter measures at the local level. Contrasting to
most previous works, \cite{janaina} proposed an approach for monitoring dengue
incidence rates based on dengue-related messages from Twitter at the city
level. The strategy consists of performing a content analysis on collected
data to assess the public perception about the disease. Based on this
analysis, a regression model is built to estimate the number of dengue cases
using the volume of messages expressing personal experience with the
disease.

Our work also focuses on estimating the dengue epidemics, a neglected tropical
disease, which is rarely explored in the context of social media channels.
Differently from most works and similarly to \cite{janaina}, our approach
considers a fine geographic-scale analysis. However, we propose a latent shared-component 
generative model that accommodates different aspects of the scenario regarding the
behavior of users in social media channels.

\section{Latent shared-component generative model}
\label{sec:method}

In this section we present in details our proposed method for estimating dengue
incidence at the city granularity.

For each town, we have two times series, $Y_t$ and $X_t$, representing both the
dengue incidence and the number of dengue-related tweets in week $t$,
respectively.  Our main hypothesis is that these two time series are positively
correlated and therefore the $X_t$ time series (dengue tweets) can provide
timely information on the dengue cases $Y_t$. We assume that the number $Y_t$
of new dengue cases in week $t$ follows a Poisson distribution with an evolving
expected value. It has been shown in \cite{Brillinger1986} that, under mild
theoretical conditions, one can expect vital statistics, such as counts of
deaths or disease cases in human populations, to follow the Poisson
distribution. This has been extensively empirically verified in epidemiological
and demographic studies to the point that the Poisson distribution is the
standard distribution to analyze these type of data.

The expected value $\mathbb{E}(Y_t)$ is equal to $\text{Pop} \times \Pi_t  ~
10^{-5}$.  That is, it is proportional to the population size and to the
dynamically evolving rate $\Pi_t$ (per 100 thousand people). This rate is a
latent stochastic process following a hidden Markov process distribution in the
form of a random walk: \[ \log(\Pi_t) = \log(\Pi_{t-1}) + \epsilon_t  \] where
the shocks $\epsilon_t$ are independent and identically distributed as a
Gaussian with mean $0$ and constant variance $1/\tau$. The reason for adopting
a random walk is to allow non-stationary movements on the dengue rate and, at
the same time, to be equally able to model stable or stationary processes. The
need to allow non-stationary movements comes from the empirical observation
that $Y_t$ present quiet weeks followed by fast increase in the number of
dengue cases, exhibiting an epidemic type of behavior.  A random walk model has
constant mean but increasing variance and naturally shows long and drastic
upward or downward movements. Hence, it is a more appropriate prior
distribution model for the time series $Y_t$ than usual stationary processes,
such as auto-regressive models, which are not able to accommodate these sudden
changes on $Y_t$.

The connection between $Y_t$ and $X_t$ is achieved through the hidden process
$\Pi_t$, a shared component that affects both counts, the dengue cases and the
dengue-related tweets in week $t$. Additionally to account for the tweets
driven by the fluctuations of the underlying Markovian rate $\Pi_t$, we allow a
stable but random amount of dengue-related tweets in each week. This additional
component acts as a random noise producing a stationary and random flow of
events on top of the disease-related tweets. We found plenty of evidence for
the presence of this random noise component when comparing the two time series
$Y_t$ and $X_t$. In several weeks where no dengue cases are reported, we may
find a few dengue-related tweets being posted. Moreover, the number of such
tweets in no-dengue weeks does not seem to vary widely or to present obvious
trends.

\begin{figure}[!htb]
\begin{center}
\includegraphics[width=1.0\linewidth]{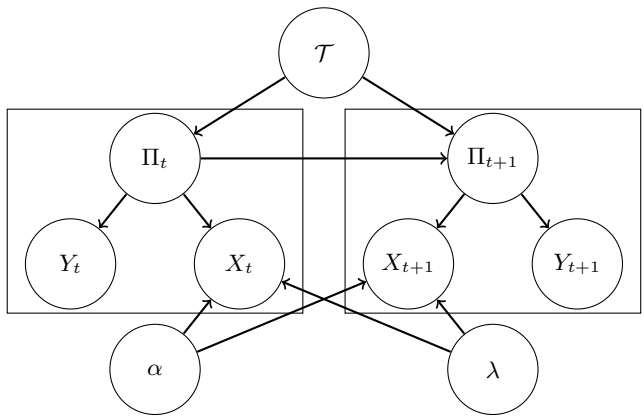}
\end{center}
\caption{Graphical model for the hidden Markov model for the time series
$Y_t$ and $X_t$ of dengue cases and dengue-related tweets in week $t$.}
\label{fig:bn}
\end{figure}

The remaining of the model assumes prior distributions for the parameters
involved in the models for $Y_t$, $X_t$, and $\Pi_t$. The graphical model
describing our Bayesian GLM model is in Figure \ref{fig:bn} and its analytical
expression with all the conditional dependencies is the following:

\begin{align}
& Y_t ~|~ \Pi_t \sim \rm{Poi}\left(\tfrac{Pop}{10^5} ~ \Pi_t \right)& \nonumber \\
& X_t ~|~ \Pi_t \sim \rm{Poi}\left(\lambda  + \tfrac{Pop}{10^5} ~ \Pi_t ~\alpha \right)& \nonumber \\
& \log(\Pi_t) = \log(\Pi_{t-1}) + \mathcal{N}\left(0, \tfrac{1}{\tau}\right)& \nonumber \\
& \lambda \sim \rm{Gamma}(0.0225, 0.0075) \\
& \alpha \sim \rm{Uniform}(0, 1) \nonumber \\
& \tau \sim \rm{Gamma}(0.01, 0.01) \nonumber
\end{align}

The model for $X_t$ is a Poisson with a basal rate $\lambda$ to capture the
random noise component on the tweets and a parameter $\alpha$ to modulate the
process $\Pi_t$, effectively bringing it to the same range of values of the
number of tweets. We selected the Gamma distribution because it is very
flexible and it is the conjugate distribution of the Poisson distribution,
facilitating the Bayesian posterior inference.  The choice for the specific
parameters of this Gamma distribution is motivated by the observation that, for
most of the towns, we know that the number of tweets in no-dengue weeks is
between 0 and 5, except for cities where the number of Twitter users is very
large and random counts in no-dengue weeks can reach up to 50 messages.  We
selected the parameters in such a way that $\lambda$ has a prior mean equal to
$0.0225/0.0075=3$ and a prior variance equal to $0.0225/0.0075^2 = 400$, which
implies in a standard deviation equals to 20 and a 1.7\% chance of being larger
than 50.

For the $\alpha$ parameter, we anticipate that the number of dengue-related
tweets will be much smaller than the number of dengue cases during the epidemic
periods. However, we do not have enough knowledge to select a narrow range for
$\alpha$ and so we adopted a flat uniform distribution in the $(0,1)$ interval.

For the variance $1/\tau$, we selected an inverse-gamma distribution (or a
Gamma distribution for the precision parameter $\tau$). Again, the reason is
mathematical convenience of its conjugacy with the normal distribution and, at
the same time, its great flexibility. The parameters $0.01$ and $0.01$ imply on
a mean equal to 1 and a standard deviation of 10. Therefore, a large range of
possible values for $\tau$ is allowable and the prior is flat enough to
accommodate itself to the observed data.

To carry out inference on the hidden structures of the model, we obtain the
posterior or conditional distribution of $\left( \{ \Pi_t \}, \lambda, \alpha,
\tau \right)$ given the data $\{ Y_t, X_t \}$.  Since analytical manipulation
of this posterior distribution is not feasible, we obtain a large sample from
it using a Markov chain Monte Carlo (MCMC) algorithm~\cite{gelman2013}.

\section{Experimental results}
\label{sec:experiments}

In this section we evaluate our latent shared-component generative model proposed in
Section~\ref{sec:method} on predicting the dengue incidence rate in Brazilian
cities using data collected from Twitter.

\subsection{Data acquisition and preprocessing}

The data used in our experiments were acquired through the Twitter streaming
application programming interface (API)~\cite{apitwitter}. In order to obtain
the dengue-related data we defined as keywords the terms ``dengue'' and
``Aedes''.  The collecting period comprises from January 2011 to December 2013.
The Twitter API provides the user's geographic location, when informed, in
three different ways: (i) geographic coordinates, in case tweets are sent from
mobile devices that release such information; (ii) an estimated location based
on the issuer IP adress; and (iii) a text field that may contain a free set of
terms provided by the user.

In Brazil the decision making process regarding public policies of dengue
surveillance are in charge of each city hall.  Therefore, we need to perform a
fine-geographic scale analysis at the city granularity. In order to do so, a
first step is to assign each message, when possible, to a valid location, or
discard it when such assignment is not possible. We start by removing from the
collected data all tweets not presenting ``Brazil'' (or ``Brasil'') in the
location field. We assume that those tweets come from places outside Brazil. We
also carefully verified the location fields removing tweets presenting Brazil
as another type of location but country. After that, we assume that all
remaining messages were posted from Brazil and we exhaustively start to seek
for consistent and non-ambiguous names of Brazilian cities along with their
respective states. Those tweets whose location were not resolved after these
steps were discarded. Table~\ref{tab:tweetsloc} provides some statistics 
about the location processing. Next we processed the messages by filtering out
accents and URL's. We also created bi-grams by joining adjacent words with a
separator, and then removed the stop-words as well as bi-grams composed of two
stop-words.

\begin{table}[!htb]
\caption{Number of tweets with respective location details}
\label{tab:tweetsloc}
\begin{tabular}{ll}
\toprule
Location details & \#tweets (\%total crawled) \\ \toprule
With geographic location & 2,096,975~(100\%) \\
``Bra[s$|$z]il'' in location field  &  1,138,199~(54.28\%) \\
``Bra[s$|$z]il'' in country field & 899,669~(42.90\%) \\ 
Location resolved & 745,617~(35.56\%) \\
\bottomrule
\end{tabular}
\end{table}

\subsection{Tweets classification and official reports}

\begin{figure}[!htb]
\begin{center}
\includegraphics[width=0.9\linewidth]{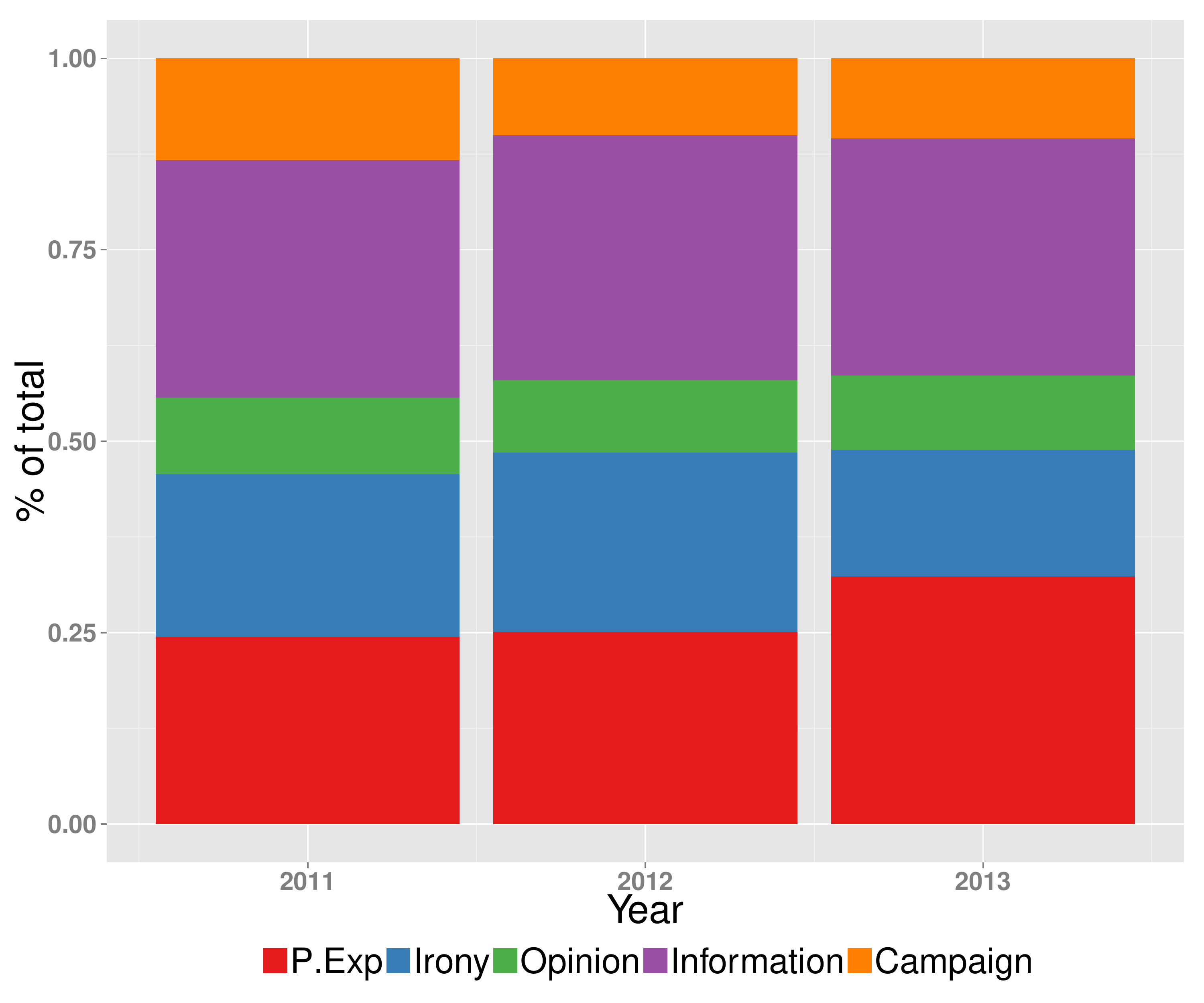}
\caption{Percentage of messages according to the category per year.}
\label{fig:bars}
\end{center}
\end{figure}

After pre-processing the tweets, we classified them in different groups
according to the sentiment they express. We performed such classification in a
supervised manner. In order to do so, we created a representative training
dataset using a selective sampling approach~\cite{sampling} and manually
labeled 2,142 tweets from 2010.  Similarly to \cite{janaina,Iberamia2014} the
tweets were classified into one out of five categories: Personal Experience,
Information, Opinion, Campaign, Irony/Sarcasm. The classification was performed
using the Lazy Associative Classification algorithm (LAC)~\cite{LAC}. This
classifier employs association rules to assign textual patterns to predefined
classes. Each rule represents a weighted vote based on the rule
confidence.  A message is then assigned to the class with the higher number of
votes. Figure~\ref{fig:bars} shows the percentage of tweets belonging to each
category over the 3 years of analysis.  Notice that the percentage of
each class remained stable through the years.

\begin{figure}[!htb]
\begin{center}
\includegraphics[width=1.0\linewidth]{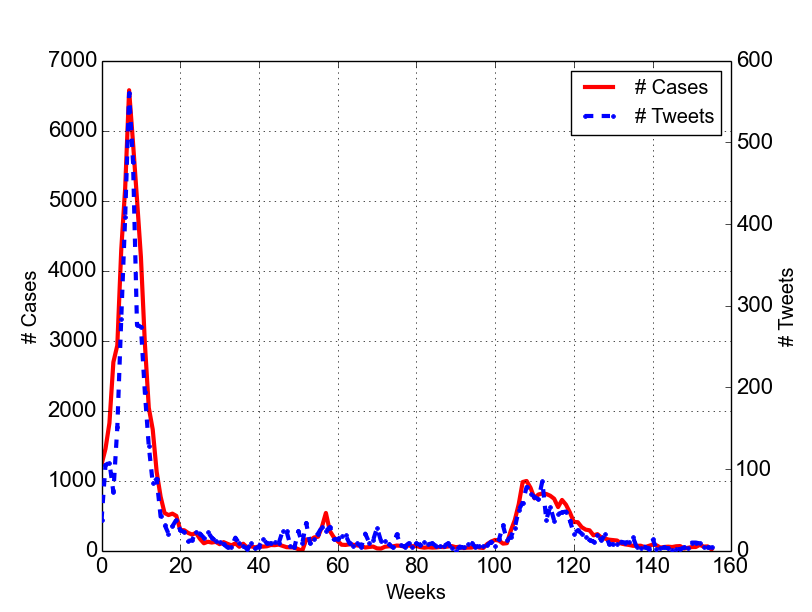}
\caption{Time series of official dengue incidence and tweets expressing personal experience for the city of Manaus.}
\label{fig:manausbh}
\end{center}
\end{figure}

The official dengue reports for each city were provided by the Brazilian Health
Ministry. The number of dengue cases is reported at a weekly granularity.
Therefore, we also aggregated the number of tweets in the same way. After
classifying and aggregating the messages, we start our analysis by observing
the correlation between the tweets and official statistics. We found that, for
most of the cities, the correlation between the number of tweets expressing
personal experience and the actual dengue cases is higher than considering the
other categories, specially for cities with a large population.  Therefore, we
decide to consider only the time series related to the tweets of that category.
Figure~\ref{fig:manausbh} shows the time series of personal experience messages
along with the official number of dengue cases for the city of Manaus, which
presents the highest correlation amongst all cities considered (0.971).

\subsection{Dengue incidence prediction}

In this section we employ our proposed model to estimate the dengue incidence
in Brazilian cities.  We restricted our analysis to the 100 cities with the
higher number of tweets. For each city, the procedure is the same, as described
next.  We fitted our Bayesian model via MCMC sampling using R in conjunction
with the OpenBUGS software~\cite{openbugs}.  We generated 2 parallel MCMC
chains, each one of length $10^6$ with a burn-in period of $500000$ and
thinning of $1000$ to obtain $1000$ samples of the joint posterior distribution.
Recall that our data is composed of two time series, $X_t$ and $Y_t$,
representing the number of tweets and official dengue reports, respectively.
The data period comprises 156 weeks, from January 2011 to December 2013. In
order to assess the prediction ability of the model, we considered a sliding
window of size 12/4, which means that we fit our model using data from 12 weeks
and estimate the number of dengue cases in the next 4 weeks only using the
previously estimated parameters and values of $X_t$ in these 4 weeks. Hence, we
have 144 prediction windows. In order to compare our model, we perform a
similar experiment using a linear univariate regression model.

We evaluate the model results using the same scale of the traditional dengue
surveillance system, defined by the World Health Organization, which comprises
three ranges (always per 100 thousand inhabitants): under 100 cases, low
incidence; 100 to 299,  medium incidence; and from 300 cases on, high incidence.
According to the estimated number of cases obtained by the model, we compute the
ranges and compare to the actual ranges obtained using the real dengue cases,
counting the number of times the actual range was predicted incorrectly.  We
assess the model first observing the range because it determines the decision
about taking control actions for the disease.

\begin{figure}[!htb]
\begin{center}
\includegraphics[width=0.9\linewidth]{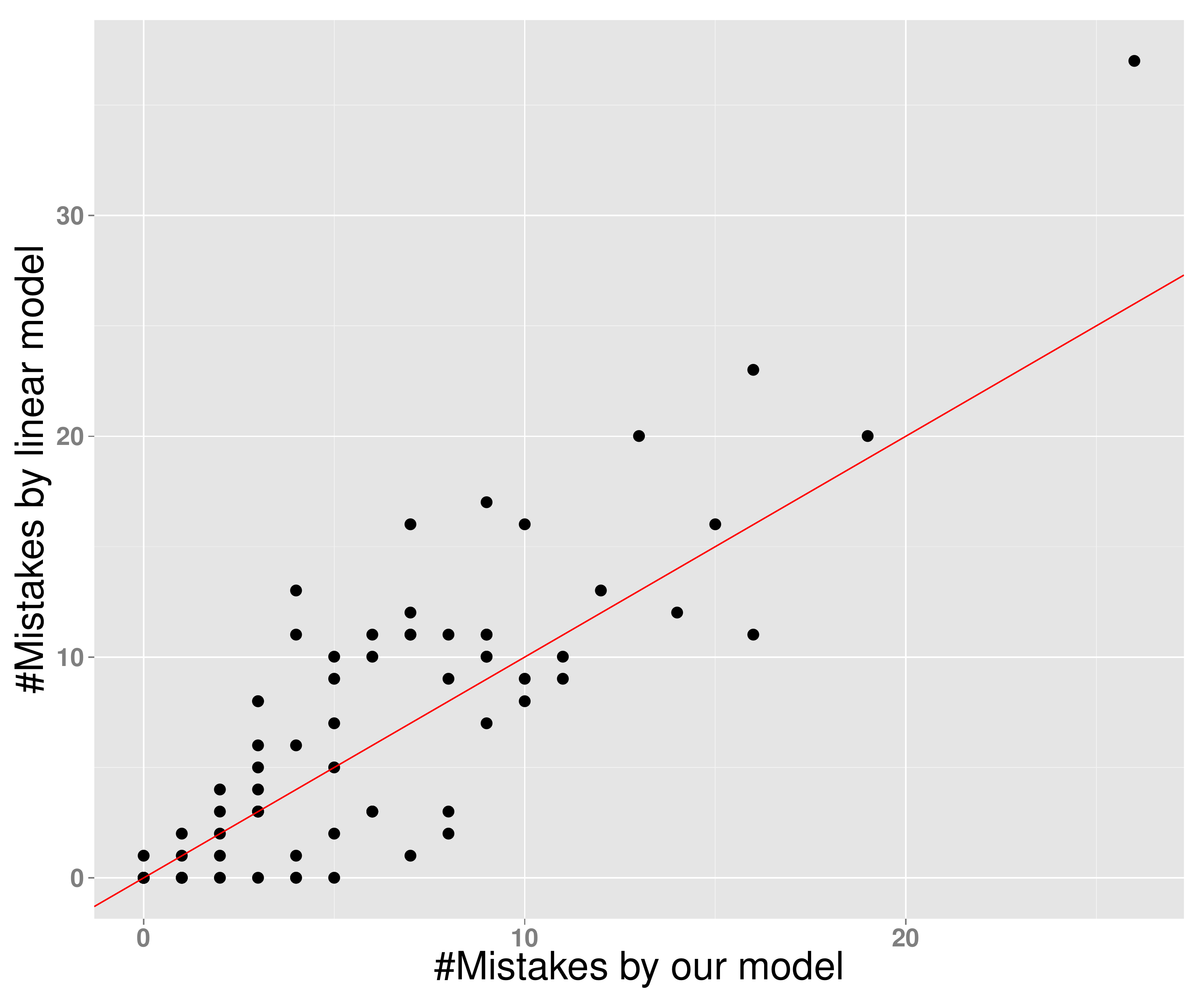}
\caption{Total number of mistakes considering the incidence scale during the sliding window process.}
\label{fig:errosfaixas}
\end{center}
\end{figure}

Figure~\ref{fig:errosfaixas} shows the results for 89 out of the 100 considered
cities.  Some cities were discarded when one or both models fail to obtain the
predictions in at least one sliding window.  This happens because during those
sliding windows, there may be a large number of zeros.

We observe that our model presents the best result for most of the cities,
performing as well as or better than the baseline in 71.9\% of the cities.
Considering wins, ties and losses we have 35, 29 and 25, respectively.  For
those cities where we observed the ties, we consider the absolute estimation
error to break the ties.  In this case, our model achieved the lower error in 19
out of the 29 ties. Thus, our model obtained the best results in 60.6\%
of the cities. The linear regression also obtained good results but we believe
that this is due to the small window size considered in the experiment. As the
window grows, the performance of the linear regression tends to degenerate
since it does not accommodate the changes in the temporal distribution of the
data. In this sense, considering a large window, our model tends to obtain
results even better than the considered baseline.

\section{Concluding remarks}
\label{sec:conclusions}

How can we take advantage of the cheap and readily available information
provided by social network users for social good?  How data mining can help?
These questions have recently intrigued researchers. In the case of developing
countries such as Brazil, health, crime and education are the issues of major
concern for the general population and they are intensively debated on social
forums. The scarce resources allocated to these areas in these countries can
receive useful and timely input from social network information. In this paper
we attempt to help on this task by proposing a simple yet effective analytic
model that puts together disease incidence and disease-related posts in a
social network. 

Our application scenario is a major infectious disease in Brazil and other
tropical countries, dengue, and we were able to estimate its incidence based on
dengue-related posts in Twitter.  The model is parsimonious in its
parameterization and requires little input from the user.  Although simple, the
model captures the main features associated with the disease process and how it
manifests in social media.  In particular, we modelled the tweets as the
superposition of two independent processes, one representing a stable emission
of a random number of posts and the other representing the additional
contribution given by the disease rate swings. We show that our model is able
to predict well the up to four weeks of the disease counts in advance using all
the available information at the time.  

We see our proposal as a first step on the direction of a more general and
comprehensive model. In fact, future research directions abound, both for
theoretical as well as experimental work. In addition to the tweets of personal
experience, we will consider the integrated use of other categories such as
tweets with dengue information, opinion, or campaigns. The model is flexible
enough to include other predictive features such as weather data. Data on
temperature and rainfall could be introduced to modulate the rate process. For
example, a $k$-dimensional feature vector $\mathbf{x}_t$ could be introduced in
the model through a link function: \[ \log(\Pi_t) = \log(\Pi_{t-1}) +
\mathbf{x}_t^{\prime} ~ \mathbf{\beta} + \mathcal{N}\left(0,
\tfrac{1}{\tau}\right) \] where the coefficients vector $\mathbf{\beta}$
captures the effect of the features in the disease rate.

A more ambitious and challenging extension is the simultaneous modeling of all
geographical units. It is reasonable to think that neighboring areas tend to
have a similar pattern on disease process because they share many of the
natural and social aspects that are relevant for the prevalence of the disease.
This could be helpful specially in small population areas that do not have
enough Twitter users to build a reliable time series of messages. The main idea
is to allow small population areas to borrow strength from the evidence
available on the neighboring areas.

\section{Acknowledgments}
The authors would like to thank INWeb, CNPq, CAPES and Fapemig for financial support.


%
\bibliographystyle{abbrv}

\begin{thebibliography}{10}

\bibitem{portalsaude}
Brazilian health ministry \\
  http://portalsaude.saude.gov.br/index.php/situacao-epidemiologica-dados-dengue.

\bibitem{who}
World health organization \\ http://www.who.int/csr/disease/dengue/denguenet.

\bibitem{Anderson1992}
R.~Anderson and R.~May.
\newblock {\em Infectious Diseases of Humans: Dynamics and Control}.
\newblock Dynamics and Control. OUP Oxford, 1992.

\bibitem{Bailey1975}
N.~Bailey.
\newblock {\em The Mathematical Theory of Infectious Diseases and its
  Applications}.
\newblock Griffin, London, 1975.

\bibitem{nature2013}
S.~Bhatt et~al.
\newblock The global distribution and burden of dengue.
\newblock {\em Nature}, 496:504--507, 2013.

\bibitem{Brillinger1986}
D.~Brillinger.
\newblock A biometrics invited paper with discussion: The natural variability
  of vital rates and associated statistics.
\newblock {\em Biometrics}, 42(4):693--734, 1986.

\bibitem{STSS}
T.~Cheng and T.~Wicks.
\newblock Event detection using twitter: A spatio-temporal approach.
\newblock {\em PLoS ONE}, 9(6):e97807, 2014.

\bibitem{Chunara}
R.~Chunara, J.~R. Andrews, and J.~S. Brownstein.
\newblock {Social and News Media Enable Estimation of Epidemiological Patterns
  Early in the 2010 Haitian Cholera Outbreak}.
\newblock {\em The American Journal of Tropical Medicine and Hygiene},
  86(1):39--45, 2012.

\bibitem{Culotta2010}
A.~Cullota.
\newblock Towards detecting influenza epidemics by analyzing twitter messages.
\newblock In {\em Proceedings of 1st Workshop on social media analytics}. ACM,
  2010.

\bibitem{gelman2013}
A.~Gelman, J.~Carlin, H.~Stern, D.~Dunson, A.~Vehtari, and D.~Rubin.
\newblock {\em Bayesian Data Analysis, Third Edition}.
\newblock Chapman \& Hall/CRC Texts in Statistical Science. Taylor \& Francis,
  2013.

\bibitem{janaina}
J.~Gomide, A.~Veloso, W.~Meira~Jr., V.~Almeida, F.~Benevenuto, F.~Ferraz, and
  M.~Teixeira.
\newblock Dengue surveillance based on a computational model of spatio-temporal
  locality of twitter.
\newblock In {\em Proceedings of the ACM WebSci Conference}, 2011.

\bibitem{sampling}
J.~Kivinen and H.~Mannila.
\newblock The power of sampling in knowledge discovery.
\newblock In {\em Proceedings of the Symposium on Principles of Databases
  Systems}, pages 77--85, 1994.

\bibitem{Lampos}
V.~Lampos and N.~Cristianini.
\newblock Tracking the flu pandemic by monitoring the social web.
\newblock In {\em Proceedings of 2nd Workshop on Cognitive Information
  Processing}. IAPR, 2010.

\bibitem{openbugs}
D.~J. Lunn, A.~Thomas, N.~Best, and D.~Spiegelhalter.
\newblock Winbugs: A bayesian modelling framework: Concepts, structure, and
  extensibility.
\newblock {\em Statistics and Computing}, 10(4):325--337, 2000.

\bibitem{Matsubara2014}
Y.~Matsubara, Y.~Sakurai, W.~G. van Panhuis, and C.~Faloutsos.
\newblock Funnel: Automatic mining of spatially coevolving epidemics.
\newblock In {\em Proceedings of the 20th ACM SIGKDD International Conference
  on Knowledge Discovery and Data Mining}, KDD '14, pages 105--114, New York,
  NY, USA, 2014. ACM.

\bibitem{clinicaldengue2013}
N.~E.~A. Murray, M.~B. Quam, and A.~Wilder-Smith.
\newblock {Epidemiology of dengue: past, present and future prospects}.
\newblock {\em Clinical Epidemiology}, 5:299--309, 2013.

\bibitem{Dredze}
M.~J. Paul and M.~Dredze.
\newblock You are what you tweet: Analyzing twitter for public health.
\newblock In {\em Proceedings ICWSM}, 2011.

\bibitem{Prieto}
V.~M. Prieto, S.~Matos, M.~Álvarez, F.~Cacheda, and J.~L. Oliveira.
\newblock Twitter: A good place to detect health conditions.
\newblock {\em PLoS ONE}, 9(1):e86191, 2014.

\bibitem{Sakaki2010}
T.~Sakaki, M.~Okazaki, and Y.~Matsuo.
\newblock Earthquake shakes twitter users: real-time event detection by social
  sensors.
\newblock In {\em Proceedings of International Conference on World Wide Web},
  pages 851--860. ACM, 2010.

\bibitem{Santos}
J.~C. Santos and S.~Matos.
\newblock Analysing twitter and web queries for flu trend prediction.
\newblock {\em Theoretical Biology and Medical Modelling}, 11, 2014.

\bibitem{Signorini}
A.~Signorini, A.~M. Segre, and P.~M. Polgreen.
\newblock The use of twitter to track levels of disease activity and public
  concern in the u.s. during the influenza a h1n1 pandemic.
\newblock {\em PLoS ONE}, 6(5):e19467, 2011.

\bibitem{Iberamia2014}
R.~C. S. N.~P. Souza, D.~E.~F. de~Brito, R.~L. Cardoso, D.~M. de~Oliveira,
  J.~Meira, Wagner, and G.~L. Pappa.
\newblock An evolutionary methodology for handling data scarcity and noise in
  monitoring real events from social media data.
\newblock In {\em 14th Ibero-American Conference on Artificial Intelligence},
  pages 295--306, 2014.

\bibitem{apitwitter}
Twitter.
\newblock The streaming api \\https://dev.twitter.com/streaming/overview.

\bibitem{LAC}
A.~Veloso, W.~M. Jr., and M.~J. Zaki.
\newblock Lazy associative classification.
\newblock In {\em Proceedings of the International Conference on Data Mining},
  pages 645--654, 2006.

\end{thebibliography}

%
%

\end{document}